\documentclass[ejsv2]{imsart}

\RequirePackage[numbers]{natbib}
\RequirePackage[colorlinks,citecolor=blue,urlcolor=blue]{hyperref}
\RequirePackage{graphicx}

\usepackage{graphicx} 
\usepackage{epstopdf}
\usepackage{float}
\usepackage{fixmath}
\usepackage{bm}
\usepackage{color}
\usepackage{enumitem}  
\usepackage{latexsym}
\usepackage{amsfonts}
\usepackage{mathtools,xparse}

\startlocaldefs
\theoremstyle{plain}

\newtheorem{theorem}{Theorem}[section]
\newtheorem{lemma}[theorem]{Lemma}

\newcommand{\eop}{\hfill $\Box$ \\ \\}

\theoremstyle{definition}

\theoremstyle{remark}

\endlocaldefs

\begin{document}

\begin{frontmatter}
\title{Estimating the distribution of marks of a homogeneous marked Poisson process}
\runtitle{Homogeneous marked Poisson process}

\begin{aug}
\author[A]{\fnms{Dragi}~\snm{Anevski}\ead[label=e1]{dragi@maths.lth.se}},
\author[B]{\fnms{Vladimir}~\snm{Pastukhov}\ead[label=e2]{vmpastukhov@yahoo.com}}
\address[A]{Centre for Mathematical Sciences, Lund University, Sweden\printead[presep={,\ }]{e1}}

\address[B]{Department of Statistics and Operations Research, University of Vienna, Austria\printead[presep={,\ }]{e2}}
\runauthor{D. Anevski and V. Pastukhov}
\end{aug}

\begin{abstract}
In this paper we propose an estimator of the distribution of events of different kinds in a homogeneous Poisson process. We give an explicit solution for the maximum likelihood estimator of the distribution and derive its strong consistency and asymptotic normality. 
We also provide an order restricted estimator of the distribution and derive its consistency and asymptotic distribution. 
The inference problem gives rise to a Sylvester-Ramanujan system of equations.
 We discuss application of the estimator to the detection of neutrons in a novel detector developed at the European Spallation Source in Lund, Sweden.
\end{abstract}



\end{frontmatter}

\section{Introduction}

The motivation for the research in this paper comes from neutron detection, and it is of importance for the European Spallation Source\footnote{\tt https://europeanspallationsource.se} (ESS) which is a large scale research facility {in Lund, Sweden}. The main research problem from a physicist's perspective  is the estimation of the energy or, equivalently,  wavelength distribution of a neutron beam. The data in the neutron scattering experiment for the neutron detector, that we are considering, consists of counts of the number of neutrons that have been absorbed along the {\em layers} in the detector. Given the data, the goal is to estimate the unknown wavelength distribution in the neutron beam that one has observed. We have previously studied this problem in a simpler setting with exactly one wavelength in the neutron beam which was then considered to be unknown, cf. \cite{anevskistocpr}. The goal in  \cite{anevskistocpr} was to derive an estimator of the unknown wavelength, which was a maximum likelihood estimator (mle), and to derive properties of the estimator. In particular, \cite{anevskistocpr} showed consistency and asymptotic normality of the mle. 

This paper can be seen as a generalisation of the study in  \cite{anevskistocpr}, in a sense that we investigate the same detector from the physicists' perspective, but we are now interested in a set of wavelengths with a finite cardinality, say $s$, and that both the wavelengths values as well as the distribution of the wavelengths in the neutron beam are unknown. The goal in this paper is to construct an estimate of these $2s$ parameters and to derive properties of the constructed estimator.  

In  \cite{anevskistocpr} the neutron beam was assumed to be well described by a time homogeneous Poisson process, and we take a similar approach here. In this paper we assume that the neutron beam is a superposition of individual neutron beams with different wavelengths each being described by a Poisson process. The proportions $\bm{q} = (q_{1}, \dots, q_{s})$ for the individual wavelengths in the total beam are, however, unknown, and $\bm{q}$ is, in fact, a parameter that we want to estimate. The total sum is, of course, still a Poisson process. The data obtained from the neutron detector consists of counts of neutrons that are absorbed and detected in the detector, and we may use the key observation that the probability of absorption of a specific neutron is, in principle,  a known function of its wavelength. Thus, each neutron in the beam will be absorbed with a probability which depends on the wavelength of that neutron, and one may assume that the absorptions of different neutrons, even of the same wavelength, are independent events. This points to the direction of modelling with the use of thinned Poisson processes. 

In fact, in this paper we treat an inference problem that can be stated as the estimation of the distribution of marks of a homogeneous marked Poisson process. Having stated the problem and formulated a maximum likelihood estimator, we see that the problem becomes difficult to treat directly if one goes trough a standard machinery of finding zeros to the score equations. In fact, the problem may be simplified by rephrasing it into estimating algebraic functions of some of the $2s$ parameters and then, having obtained the estimates of the algebraic functions,  trying to solve the upcoming algebraic equations for the variables in those equations. This later problem can be seen as a problem of solving a system of $k$ algebraic equations for $(\bm{q},\bm{p})\in{\mathbb R}^{2s}$, which in our setting can be written as
\begin{eqnarray}\label{sylrameqs}
     \sum_{r=1}^s (1- p_{r})p_{r}^{i-1}q_{r}&=&  \hat{b}_n^{(i)},
\end{eqnarray}
for $i=1,\ldots,k$, where  $\hat{b}_n^{(i)}$ are given statistics, cf. (\ref{fmi}) below. 

The system of equations (\ref{sylrameqs}) was first studied by Sylvester in \cite{sylvester:1851}, then by Ramanujan in \cite{Raman} and with later refinements given in \cite{lyubich:2004}. Following \cite{lyubich:2004}, we call the system (\ref{sylrameqs}) the Sylvester-Ramanujan system.
In our setting $\bm{q}$ denotes the distribution of marks (i.e. the wavelengths in the neutron beam), while $\bm{p}$ denotes the thinning probabilities for the respective wavelengths. 
As shown by Ramanujan, a sufficient number of equations to obtain a solution is $2s-1$, and thus $k$ above should be $2s-1$. 

The Sylvester-Ramanujan system  is also related to the Hamburger power moment problem (see, e.g. \cite{lyubich:2004, prekota:2004} with the references therein) which consists of finding the distribution of a random variable if one knows all power moments, which for a finite support case can be written as the problem of finding the probability mass function $\bm{f}=(f_{1}, \ldots, f_{s})$ that solves
\begin{eqnarray*}
   \sum_{i=1}^s i^j f_{i}&=&m_j,
\end{eqnarray*}
for given $m_j$, with $j=0,\ldots,2s-1$.

Using standard results on almost sure consistency and asymptotic normality for the mle, coupled with continuity and differentiability of the function that defines the solution of the Ramanujan equations, via the continuous mapping theorem and the delta method we obtain almost sure consistency and asymptotic normality of the desired mle of $(\bm{q},\bm{p})$. Taking into account the fact that often the set of frequencies in a beam is a basic frequency and its overtones or, equivalently, that the set of wavelengths consists of a dominant wavelength and its fractions, it makes sense to model the wavelength distribution $\bm{q}$ as a decreasing sequence. This is a motivation for finding an order restricted estimator of $\bm{q}$, and we therefore propose the $l^2$-projection of the unrestricted mle of $\bm{q}$ on the set of decreasing probability mass functions. We are then able to use results on consistency and limit distribution  for such isotonic regression estimators, see \cite{jankowski:wellner:2009, robertsonorder} for the results for i.i.d data and \cite{anevski:pastukhov:2018-general} for general results.

The remainder of the paper is organised as follows. In Section \ref{sec:motivation} we give a detailed description of the detector model that is being used, Poisson process model for the beam and the data generated from the detector, cf. Lemma \ref{indepcounts}. In Section \ref{sec:inference-parameters} we study the likelihood approach for estimation of the parameters $(\bm{q},\bm{p})$ and the system of algebraic equations that facilitates the estimation. In Theorem \ref{qptild} we show that if the number of equations is $k=2s-1$, then there is a unique mle of  $(\bm{q},\bm{p})$ obtained by the solution of the algebraic equations. In Theorem \ref{thmmle} we derive consistency and asymptotic normality of the mle of  $(\bm{q},\bm{p})$. In Section \ref{sec:order-restricted} we define an order restricted estimator of $\bm{q}$ and state its consistency and asymptotic distribution in Theorem \ref{thm:order-mle}. Finally, in Section \ref{sec:discussion} we discuss the obtained results and some remaining and interesting future problems.

\section{Motivation and description of the data generating mechanism}\label{sec:motivation}
The inference problem is motivated by the following problem that arises in neutron detection. Assume that a neutron beam comes to the face of the detector. We model the number of neutrons that arrives in the time interval $[0,t]$ by a counting process $X_0(t)$. Assume that the neutron beam, i.e. the process $X_0(t)$, has a constant intensity $\lambda$. Assume furthermore that there are  $s > 1$ different kinds of neutrons in the beam with different wavelengths $\bm{\mu} = (\mu_{1}, \dots, \mu_{s})$, such that
\begin{eqnarray}\label{eq:ordered-wavelengths}
\mu_{1} < \mu_{2} < ... < \mu_{s}.
\end{eqnarray}
The values of the wavelengths are unknown. 

We model the total neutron beam, the counting process $X_0(t)$, as the sum of counting processes of the individual types  neutrons that count the number of neutrons that arrives into detector in the time interval $[0,t]$. Thus, we let the number of neutrons with the wavelength $\mu_{r}$, which we may label as $r$-neutrons, be denoted by $X^{(r)}_{0}(t)$, where $X^{(r)}_{0}(t)$ is a  counting process such that $X^{(r)}_0(t) =0$ and with intensity $\lambda_r$, for $r = 1, \dots, s$. 
Then, $X_0(t) = \sum_{r=1}^s X^{(r)}_{0}(t)$ is the total number of neutrons that arrives into the detector.

For a given total number of incoming neutrons in the time interval $[0,t]$, say $X_0(t)=x_{0} $, the vector $\big(X^{(1)}_{0}(t), X^{(2)}_{0}(t), ..., X^{(s)}_{0}(t)\big)$ is assumed to follow a multinomial distribution with parameters $\big(  q_{1}, q_{2}, ...,q_{s}\big)$,  i.e.

\begin{eqnarray}\label{multinX}
(X_{0}^{(1)} = x_{0}^{(1)}, ..., X_{0}^{(s)} = x_{0}^{(s)} | X_{0} = x_{0}) &\in& Mult(x_{0},  q_{1}, q_{2}, ...,q_{s}),
\end{eqnarray}
with

\begin{eqnarray*}\label{}
x_{0}^{(1)} +x_{0}^{(2)}+ \dots +x_{0}^{(s)} &=& x_{0},\\
q_{1} + q_{2} + ... + q_{s} &=& 1.
\end{eqnarray*}
The vector of proportions of the numbers of different neutrons $\bm{q} = (q_{1}, q_{2}, ...,q_{s})$ is the spectrum, or distribution, of an incoming neutron beam $X_0(t)$. We assume that $\bm{q}$ does not depend on $t$.

Now assume that the incident beam $X_{0}(t)$ is a Poisson process with an intensity $\lambda$.
In this case the components $X^{(r)}_{0}(t)$, for $r=1,\ldots,s$, of the beam are independent Poisson processes with intensities $\lambda_r = q_{r}\lambda$, for $r = 1, \dots, s$, since the vector  $\big(X^{(1)}_{0}(t), X^{(2)}_{0}(t), ..., X^{(s)}_{0}(t)\big)$ is the thinning of the original Poisson process, e.g. cf. \cite{Assuncao}. 
 
\begin{figure}[htbp!]
\center
\includegraphics[width=1\linewidth]{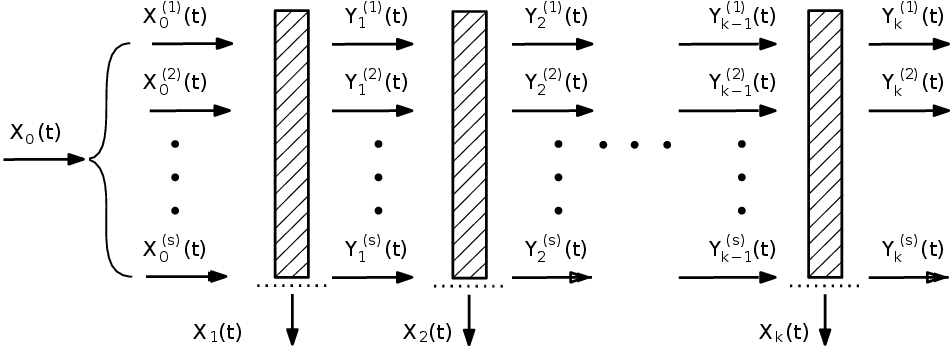}
\caption{The scheme of the detector. Here $Y^{(r)}_{i}(t)$ is the number of transmitted  $r$-neutrons and $X_{i}(t) = \sum_{r=1}^s X^{(r)}_{i}(t)$ is the total number of the neutrons absorbed at the layer $i$.}\label{detektorp0}
\end{figure}

Next, we introduce the so called multilayer detector that is used in this setting. The detector consists of a fixed number of layers, say $k > 1$ layers, as displayed in Fig. \ref{detektorp0}. The value of $k$ will be elaborated on below and it will be shown to be determined by the number of different types of neutrons that are present in the neutron beam.

The detection of neutrons in the multilayer detector can be described as follows. When an incident beam of neutrons hits a layer of the detector, each neutron in the beam can possibly be absorbed and then detected or, otherwise, not absorbed. If the neutron is not absorbed it will go through the present layer and will subsequently arrive at the next layer. We assume that at each layer absorption and transmission  are the only possibilities for the neutron interactions with the layer. We also assume that at each layer different particles interact with the  layer independently of each other, i.e. at each layer the absorptions of different neutrons are independent events. 

Let $\bm{p} = (p_{1}, \dots, p_{s})$ be the vector of probabilities of transmission (the thinning parameters), so that $1-p_{r}$ is the probability of absorption for $r$-neutron, for $r = 1, \dots, s$. It is a physical property of a neutron that the probability of transmission decreases with the neutron wavelength, cf. \cite{AnevskiWilton} and references therein, and, therefore, the thinning parameters can be modelled as a decreasing sequence 
\begin{eqnarray}\label{tpp}
1 > p_{1} > p_{2} > ... > p_{s} > 0.
\end{eqnarray}

Let us consider a beam of $r$-neutrons and denote the number of $r$-neutrons that are absorbed at the first layer by $X_{1}^{(r)}(t)$ so that $Y^{(r)}_{1}(t)  =  X^{(r)}_0(t)-X^{(r)}_{1}(t)$ is the number of $r$-neutrons that are transmitted. Then, $X^{(r)}_{1}(t)$ and $Y^{(r)}_{1}(t) = X^{(r)}_0(t) - X^{(r)}_{1}(t)$ are non-decreasing counting processes obtained by thinning of the original Poisson process $X^{(r)}_0(t)$ so that $X^{(r)}_{1}(t)$ and $Y^{(r)}_{1}(t)$ are independent Poisson processes with intensities $(1 - p_{r})q_{r}\lambda$ and $p_{r}q_{r}\lambda$, respectively, cf. \cite{Assuncao}.

Now assume that the transmitted beam $Y^{(r)}_{1}(t)$ hits the second layer, at which, again, each $r$-neutron can be  either absorbed or transmitted. Let $X^{(r)}_{2}(t)$ be the number of absorbed neutrons and $Y^{(r)}_{2}(t)=Y^{(r)}_{1}(t) - X^{(r)}_{2}(t)$ the number of transmitted neutrons at the second layer. Then, again, $X^{(r)}_{2}(t)$ and $ Y^{(r)}_{2}(t)$ are obtained by the thinning of the Poisson process $Y^{(r)}_{1}(t)$ and, therefore, they  are independent Poisson processes with intensities  $p_{r}(1-p_{r})q_{r}\lambda$ and $p_{r}p_{r}q_{r}\lambda$, respectively, cf. \cite{Assuncao}. By iterating the argument, cf. also \cite{anevskistocpr} for a similar and more detailed reasoning, we obtain the following result.
\begin{lemma}\label{indepcounts}
$\{X_{i}(t)\}$, for $i = 1, \dots, k $, are jointly independent Poisson processes with intensities $\sum_{r=1}^s(1- p_{r})p_{r}^{i-1}q_{r}\lambda$, respectively.
\end{lemma}

The goal of this paper is the estimation of the wavelength distribution $\bm{q}$ of the incident beam as well as of the actual values of the wavelengths $\bm{\mu}$, based on observations of the (total) Poisson process and with the use of the multilayer neutron detector, described above.  Estimators of the wavelength values  $\bm{\mu}$ can be indirectly obtained via estimates of the thinning parameters $\bm{p}$, using a functional relation between the wavelength and the thinning probability as explained in \cite{anevskistocpr}. {One may ask whether the wavelength distribution  $\bm{q}$ can be left completely arbitrary or whether is should satisfy some restriction other than being positive and summing to one. A reasonable model to impose arises from the argument from physics that one may assume that the lowest frequency is the most significant in the beam and that there is a subsequentially smaller amount of overtone frequencies,  cf. e.g. \cite{AnevskiWilton}. This suggests to model $\bm{q}$ as an decreasing sequence. Thus if the wavelength values are denoted so that (\ref{eq:ordered-wavelengths}) holds then one may impose the model assumption that  $\bm{q}$ is decreasing. i.e. that
  \begin{eqnarray} \label{eq:decreasing-distribution}
 q_{1} \geq q_{2} \geq  \dots  \geq q_{s}\geq 0.   
\end{eqnarray}
}

{ In light of the order restriction (\ref{eq:ordered-wavelengths}), which is equivalent to (\ref{tpp}),  and (\ref{eq:decreasing-distribution}) we note that the inference problem can be seen as simultaneously estimating the wavelengths and distribution  $(\bm{\mu},\bm{q})$ under the order restrictions (\ref{eq:ordered-wavelengths}) and (\ref{eq:decreasing-distribution}), or equivalently simultaneously the transmission probablities and distribution $(\bm{p},\bm{q})$ under the order restrictions  (\ref{tpp}) and (\ref{eq:decreasing-distribution}).}

{
  We first note a special feature in the data that we obtain in this setting, namely that the actual order of the wavelengths is not observed. This is a result of the fact that the observed number of neutrons $X_i(t)$ are sums of the $s$ number of neutrons that have {\em one fixed wavelength} $\lambda_r$ for $r=1,\ldots,s$, and it turns out that we can not separate the different wavelength neutron counts. A further feature in our inference problem is that the parameters are jointly ordered, i.e. they satisfy  (\ref{tpp}) and (\ref{eq:decreasing-distribution}) as well as $\sum_{r=1}^s q_r=1$. The two order restrictions are equivalent to a partial order $\succeq$  on ${\mathbb R}^2$  defined by
  \begin{eqnarray*}
   (x_1,y_1) \succeq (x_2,y_2)&\Leftrightarrow& x_1 > x_2 \; \cap \; y_1\geq y_2,
  \end{eqnarray*}
one should note the strict inequality on the first coordinate. With this partial order we can express the order assumptions (\ref{tpp}) and (\ref{eq:decreasing-distribution}) equivalently as
\begin{eqnarray}\label{eq:partial-order}
  (p_1,q_1)\succeq (p_2,q_2)\succeq \ldots \succeq (p_r,q_r)\succeq (0,0).
  \end{eqnarray}
}

{
Thus we have the inference problem of estimating the pairs $(p_r,q_r)$ for  $r=1,\ldots,s$ subject to the (partial) order restriction (\ref{eq:partial-order}) and $\sum_{r=1}^sq_r =1$, when {\em the labels are unknown}. We can explain in detail the last assumption as follows: We suppose that we are given $(\hat{p}_1,\hat{q}_1),\ldots,(\hat{p}_s,\hat{q})_s$ as some preliminary estimator of the $s$ pairs, that may or may not satisfy the order relation (\ref{eq:partial-order}). It is crucial however that we are {\em not given the preliminary estimator indexed by the correct labels}; rather we suppose that we are merely given the $s$ pairs as elements in an unordered set. A one dimensional order restricted inference problem with unknown labels, that originates in computer science applications and missing species problems, is treated in \cite{anevski:gill:zohren:2017}.  
}

{
  We see that the parameter space defined by (\ref{eq:partial-order}) and $\sum_{r=1}^sq_r =1$ is not closed, and thus an optimisation approach, such as a likelihood based estimator, over that parameter space typically will not guarantee a solution that is in the correct parameter space. We therefore refrain from using such an order restricted approach for the maximum likelihood problem. Instead we formulate a likelihood problem, that is not order restricted, in  a way that will give us the Hamburger moment problem, and are able to use the Ramanujan equations to find solutions to the maximum likelihood problem as pairs $(\hat{p}_1,\hat{q}_1),\ldots,(\hat{p}_s,\hat{q}_s)$ such that the estimated $\hat{p}_r$ values, $r=1,\ldots,s$, are all distinct. One may then attempt to use the values  $(\hat{p}_r,\hat{q}_r)$, $r=1,\ldots,s$ as a preliminary estimator and attempt to do bivariate isotonic regression, cf. \cite{anevski:pastukhov:2018-general} for a definition and the asymptotic distribution of such an order restricted estimator. This attempt may however lead to estimated values $\hat{p}_r$ that will be the same for {\em different} wavelengths, i.e. one may have $\hat{p}_i=\hat{p}_j$ for $i\neq j$, which is a physical impossibility.
}

{
Thus we will not use this seemingly more direct approach. Instead we will estimate the correct order by ordering the estimated $\hat{p}_r$ values arising from the solution to the Ramanujan equation, resulting in say $\hat{\bm p}_{\hat{\xi}}=(\hat{p}_{\hat{\xi}(1)},\ldots,\hat{p}_{\hat{\xi}(r)})$ as the ordered vector, with $\hat{\xi}$ a permutation that is estimated by the data and thus is random.  The resulting estimators $\hat{q}_r$, $r=1,\ldots,s$, we may then order as   $\hat{\bm q}_{\hat{\xi}}=(\hat{q}_{\hat{\xi}(1)},\ldots,\hat{q}_{\hat{\xi}(r)})$, since $\hat{q}_r$ should be paired with the corresponding $\hat{p}_r$. Finally we treat the vector $\hat{\bm q}_{\hat{\xi}}$  as a preliminary estimator and do a one dimensional isotonic regression of the preliminary estimator  $\hat{\bm q}_{\hat{\xi}}$, to obtain an ordered probability vector $\tilde{\bm q}$. This will give us the final estimated transmission probabilities and wavelength distribution $(\hat{\bm p}_{\hat{\xi}},\tilde{\bm q})$, which will satisfy the order restriction (\ref{eq:partial-order}) and for which we are able to derive limit distributions.
  }

\section{The  maximum likelihood estimator of $(q,p)$}\label{sec:inference-parameters}

{In this section we suggest a likelihood approach for the inference problem without the order restrictions  (\ref{tpp}) and (\ref{eq:decreasing-distribution})}, define the mle of the parameters $(\bm{q}, \bm{p})$, state conditions for its existence, and derive consistency and asymptotic normality for the mle of $(\bm{q},\bm{p})$.

We start by the following note on the experimental setup and the data. In order to derive the limit properties for the estimator, we need to define what we mean by ``letting $n$ go to infinity''. This may be done in, at least, two ways. We can either let the time $t$ go to infinity and view the data as stemming from a Poisson process which is run for a (very) long time, or we can keep the time $t$ fixed and gather data from several independent Poisson process runs, cf. \cite{anevskistocpr} for a more detailed discussion about advantages and disadvantages with respective approach.

We will choose the second approach and view the estimation problem as a repeated sample problem. Thus, we assume that during a fixed time interval $[0, t]$ and for a fixed intensity $\lambda$, i.e. fixed intensities $(\lambda_1,\ldots,\lambda_s)$, of an incident beam $X_0(t)$ there are $n$ repeated measurements. Let $x_{i, j}$ be the observed number of neutrons at layer $i$, for $i=1,\ldots,k$, at the experiment round $j$, for $j=1,\ldots,n$. Then, at each experiment round $j$ the vector $\mathbf{X_j} = (X_{1j},\ldots,X_{kj})$ is distributed according to Lemma \ref{indepcounts}, and, furthermore, the vectors $\mathbf{X_1},\ldots, \mathbf{X_n}$ are independent.

Given the data as above, the inference problem is the estimation of the pair $(\bm{q,p})$ under the restriction that they lie in the parameter space $\mathcal{F} \subset \mathbb{R}^{2s}_+$, which is given by {
\begin{eqnarray}\label{pqconstr} 
  \mathcal{F}& =& \{(\bm{q,p})\in{\mathbb R}^{2s}_+:  q_{1} + q_{2} +  \dots  + q_{s} = 1, \nonumber\\
 && 0\leq p_1,p_2,\ldots,p_s\leq 1\}.
\end{eqnarray}
}Note that $\bm{q}$ is a probability mass function while $\bm{p}$ is merely a vector of the thinning probabilities. {Note that $\mathcal{F}$ is a closed set.} We would like to emphasise here that the main object of study is the wavelength distribution $\bm{q}$. However, the thinning probabilities $\bm{p}$ are also of interest, since they determine the values of the wavelengths which, we assume, are unknown. If we know the actual wavelengths values, there is no need to estimate the thinning probabilities. {Note furthermore that we do not know the order of the wavelength values, in the indices that we use, cf. the discussion in the last paragraph of Section \ref{sec:motivation} and in Section \ref{sec:discussion}.}

We will use the likelihood approach for making inference about the unknown parameters $({\bm{q}},{\bm{p}})$. {Thus} we define the mle of $({\bm{q}},{\bm{p}})$ by
\begin{eqnarray}\label{mle1}
(\hat{\bm{q}}_{n}, \hat{\bm{p}}_{n}) &=&  \underset{(\bm{q}, \bm{p}) \in \mathcal{F}}{\arg \max}  \, l_{n}(\bm{q}, \bm{p}),
\end{eqnarray}
where
\begin{eqnarray}\label{loglikehn}
     l_{n}(\bm{q}, \bm{p}) = \sum_{j=1}^{n} \sum_{i=1}^{k}( -    \lambda t m_{i} + x_{i,j}\log m_{i} + x_{i,j}\log (\lambda t)  - \log x_{i,j}! )
 \end{eqnarray}
is the log likelihood, and
\begin{eqnarray}\label{eq:mi-def}
  m_i&=&m_i(\bm{q,p})\nonumber \\
  &=&\sum_{r=1}^s (1- p_{r})p_{r}^{i-1}q_{r}
\end{eqnarray}
 is the total expected number of absorbed neutrons at layer $i$ divided by the intensity $\lambda$ and the time $t$. 


\subsection{{The moment problem and uniqueness of the unrestricted mle}}
{In this subsection we will simplify the inference problem in order to obtain an explicit expression for the mle, defined in $(\ref{mle1})$.}


{Note first that since ${\mathcal F}$ is closed, the mle $(\hat{\bm{q}}_{n},\hat{\bm{p}}_{n})$ exists.}  {One can show} that the log-likelihood $l_{n}(\bm{q}, \bm{p} \vert \mathbf{x})$, {defined in $(\ref{loglikehn})$,} seen as a function on the parameter space  $\mathcal{F} \subset \mathbb{R}^{2s}$ is not a concave function which makes it difficult to find a solution $(\hat{\bm{q}}_{n}, \hat{\bm{p}}_{n})$ even numerically. We will therefore reparametrise the problem as an inference problem for the vector $(m_{1}, \dots, m_{k})$ of expected total numbers of  observed neutrons divided by $\lambda t$, {defined in $(\ref{eq:mi-def})$}. {Having obtained an estimator of $(m_{1}, \dots, m_{k})$ for this simpler inference problem, we will next solve an upcoming system of equations in the unknown $({\bm{q}},{\bm{p}})$} and finally obtain a solution to (\ref{mle1}). 

Introduce the notation $\hat{\bm{b}}_{n} = (\hat{b}_{n}^{(1)}, \dots, \hat{b}_{n}^{(k)})$, where
\begin{eqnarray*}\label{}
\hat{b}_{n}^{(i)} &=& \frac{\sum_{j=1}^{n}x_{i,j}}{n \lambda t}.
\end{eqnarray*}
We then rewrite (\ref{loglikehn}) as
\begin{eqnarray}\label{fmi}
  g(m_{1}, \dots, m_{k})& :=& \sum_{i=1}^k (-m_i + \hat{b}_{n}^{(i)}\log m_{i}) \\
 &=& \frac{l_{n}(\bm{q}, \bm{p} \vert \mathbf{x})}{n \lambda t}\nonumber
 \end{eqnarray}
and note that we have dropped the  last two terms in (\ref{loglikehn}) in the last equality. The function $g(m_{1}, \dots, m_{k})$ {is concave and} reaches its unique global maximum at $\hat{m}_i=\hat{b}_{n}^{(i)}$, for $i = 1, \dots, k$. Therefore, if $(\tilde{\bm{q}}_{n}, \tilde{\bm{p}}_{n})$ is a solution of the following system of equations
\begin{eqnarray}\label{fmi}
\left\{ 
\begin{array}{c}
 m_{1}(\bm{q}, \bm{p}) = \hat{b}^{(1)}_{n}\\ 
m_{2}(\bm{q}, \bm{p}) = \hat{b}^{(2)}_{n}\\ 
\dots \\
m_{k}(\bm{q}, \bm{p}) = \hat{b}^{(k)}_{n},
\end{array}
\right. 
\end{eqnarray}
where $m_i(\bm{q,p})$ are defined in (\ref{eq:mi-def}), and if it satisfies the constraints in (\ref{pqconstr}), i.e. $(\tilde{\bm{q}}_{n}, \tilde{\bm{p}}_{n}) \in \mathcal{F}$, then $(\hat{\bm{q}}_{n}, \hat{\bm{p}}_{n}) = (\tilde{\bm{q}}_{n}, \tilde{\bm{p}}_{n})$, i.e. the solution of (\ref{fmi}) is {an} mle of $(\bm{q,p})$. {We note that in that case  $ (\tilde{\bm{q}}_{n}, \tilde{\bm{p}}_{n})$ is not necessarily the unique mle.}

{We next reformulate the system of equations (\ref{fmi}) on} matrix form. We introduce the vector $\bm{\hat{a}}_{n} = (\hat{a}_{n}^{(1)}, \dots,  \hat{a}_{n}^{(k+1)})$, where 
\begin{eqnarray}\label{eq:anhat-def}
\hat{a}_{n}^{(1)} &=& 1,\nonumber\\
 \hat{a}_{n}^{(i)} &=& 1 - \sum_{l=1}^{i-1} \hat{b}_{n}^{(l)},
\end{eqnarray}
{as well as (it's expectation) the vector $\bm{{a}} = ({a}^{(1)}, \dots, {a}^{(k+1)})$, given by
\begin{eqnarray}\label{eq:a-def}
{a}^{(1)} &=& 1,\nonumber\\
{a}^{(i)} &=& 1 - \sum_{l=1}^{i-1}  m_{l}(\bm{q}, \bm{p}),
\end{eqnarray}
for $i=2,\ldots,k+1$, with $m_{l}(\bm{q}, \bm{p})$ defined in  (\ref{eq:mi-def}). Define} furthermore the matrices $\bm{C}(\bm{u})$ and $\bm{D}(\bm{u})$, as 

\begin{eqnarray}\label{Chn}
\bm{C}(\bm{u}) = 
 \begin{pmatrix}
  u_{s} & u_{s-1} & u_{s-2} & \cdots & u_{1} \\
  u_{s+1}  & u_{s}  & u_{s-1}  &  \cdots & u_{2}  \\
  u_{s+2}  & u_{s+1}  & u_{s}  &  \cdots & u_{3}  \\
  \vdots  & \vdots  & \vdots &  \ddots  & \vdots  \\
  u_{2s-1} & u_{2s-2} & u_{2s-3}& \cdots & u_{s}
 \end{pmatrix}
\end{eqnarray}
and 
\begin{eqnarray}\label{Dhn}
\bm{D}(\bm{u}) = 
 \begin{pmatrix}
  0 & 0 & 0 & \cdots & 0 \\
  u_{1}  & 0  & 0  &  \cdots & 0 \\
  u_{2}  & u_{1} & 0  &  \cdots & 0 \\
  \vdots  & \vdots  & \vdots &  \ddots  & \vdots  \\
  u_{s-1} & u_{s-2} & u_{s-3} & \cdots & u_{1}  
 \end{pmatrix},
\end{eqnarray}
for $\bm{u} \in \mathbb{R}^{2s}$.


Next, we study the system of equations (\ref{fmi}). We will follow closely Ramanujan's derivation of the solution in \cite{Raman}. Thus we define the function 
\begin{eqnarray}\label{eq:phi-def}
\varphi(\theta) = \frac{d_{1} + d_{2} \theta + d_{3} \theta^{2} + \dots + d_{s} \theta^{s-1}}{1 + c_{1} \theta + c_{2} \theta^{2} + \dots + c_{s} \theta^{s}},
\end{eqnarray}
with the vectors $\bm{c}=(c_1,\ldots,c_s)$ and  $\bm{d}=(d_1,\ldots,d_s)$ given by
\begin{eqnarray*}\label{}
\bm{c} &=&  \bm{C}(\bm{\hat{a}}_{n})^{-1} [\bm{\hat{a}}_{n}]^{(s+1, 2s)},\\
\bm{d} &=& [\bm{\hat{a}}_{n}]^{(1, s)} + \bm{D}(\bm{\hat{a}}_{n}) [\bm{c}]^{(1,s-1)},
\end{eqnarray*}
and where $[\bm{\hat{a}}_{n}]^{(i, j)} \in \mathbb{R}^{j-i+1}$ denotes the restriction of the vector $\bm{\hat{a}}_{n}$ in $\mathbb{R}^{k+1}$ to the index set $(i, j)$.

The next result says that if {there is} a certain relation between the number of layers and the support of the wavelength distribution, then the mle exists {eventually, almost surely}, and it is unique up to permutations of the indices.

\begin{theorem}\label{qptild}
{Assume that $k=2s-1$. Let $\phi$ be the function defined in $(\ref{eq:phi-def})$. Then the equations (\ref{fmi}) have a solution eventually, almost surely. When the solution exists, it is unique up to permutations of the indices. Furthermore, when the solution exists, it is given by
\begin{eqnarray*}\label{}
(\tilde{\bm{q}}_{n}, \tilde{\bm{p}}_{n}) = (\bm{y}, \bm{z}),
\end{eqnarray*}
where  $\bm{y}, \bm{z} \in \mathbb{R}^{s}$ are the coefficients in the following representation of  $\varphi(\theta)$
\begin{eqnarray}\label{phyz}
\varphi(\theta) = \frac{y_{1}}{1 -z_{1}\theta} + \frac{y_{2}}{1 -z_{2}\theta} + \dots + \frac{y_{s}}{1 -z_{s}\theta}. 
\end{eqnarray}
Moreover, when the solution exists, the components of $\tilde{\bm{p}}_{n}$ are pairwise different.}
\end{theorem}
\textbf{Proof}
Note first that the matrix $C:=\bm{C}(\bm{a})$ is given by
\begin{eqnarray}\label{mC}
\bm{C} = 
 \begin{pmatrix}
  \sum_{r=1}^{s}p_{r}^{s-1}q_{r} &\sum_{r=1}^{s}p_{r}^{s-2}q_{r} &\sum_{r=1}^{s}p_{r}^{s-3}q_{r} & \cdots & \sum_{r=1}^{s}q_{r}  \\
  \sum_{r=1}^{s}p_{r}^{s}q_{r}  & \sum_{r=1}^{s}p_{r}^{s-1}q_{r}  & \sum_{r=1}^{s}p_{r}^{s-2}q_{r}  &  \cdots & \sum_{r=1}^{s}p_{r}q_{r}   \\
  \sum_{r=1}^{s}p_{r}^{s+1}q_{r} &  \sum_{r=1}^{s}p_{r}^{s}q_{r}  &  \sum_{r=1}^{s}p_{r}^{s-1}q_{r}  &  \cdots &  \sum_{r=1}^{s}p_{r}^{2}q_{r}  \\
  \vdots  & \vdots  & \vdots &  \ddots  & \vdots  \\
  \sum_{r=1}^{s}p_{r}^{2s -2}q_{r} & \sum_{r=1}^{s}p_{r}^{2s -3}q_{r} &  {\sum_{r=1}^{s}p_{r}^{2s -4}q_{r} } & \cdots & \sum_{r=1}^{s}p_{r}^{s - 1}q_{r} 
 \end{pmatrix}.
\end{eqnarray}

Furthermore, note that {$\bm{C}(\bm{{a}})$} can be diagonalized as
\begin{eqnarray*}\label{}
 {\bm{C}(\bm{{a}})} &=&\bm{V}\bm{Q}\bm{V}^{T},
\end{eqnarray*}
where
\begin{eqnarray*}\label{}
\bm{V} = 
 \begin{pmatrix}
 1 & 1 & 1 & \cdots & 1  \\
  p_{1}  & p_{2}  & p_{3}  &  \cdots & p_{s}  \\
  p_{1}^{2} & p_{2}^{2} & p_{3}^{2} &  \cdots & p_{s}^{2} \\
  \vdots  & \vdots  & \vdots &  \ddots  & \vdots  \\
  p_{1}^{s-1} & p_{2}^{s-1} & p_{3}^{s-1} & \cdots & p_{s}^{s-1} 
 \end{pmatrix}
\end{eqnarray*}
and $\bm{Q}$ is a diagonal matrix {with $(\bm{Q})_{i,i}=q_i$}. Since $\bm{V}$ is a square Vandermonde matrix and  { since by assumption $p_{1} > p_{2} > ... > p_{s}$, cf. (\ref{tpp})}, {$\bm{V}$ is full-rank, which implies that} $rank(C) = s$. Therefore {$\det(\bm{C}(\bm{{a}})) \neq 0$}.

{From the strong law of large numbers one has
\begin{eqnarray} \label{eq:an-lim-as}
     \hat{\bm{a}}_{n} &\stackrel{a.s.}{\to}& \bm{a},
  \end{eqnarray}
cf. (\ref{eq:anhat-def}) and (\ref{eq:a-def}), and thus by the continuous mapping theorem,}
\begin{eqnarray*}\label{eq:limit-of-cn}
\bm{C}(\bm{\hat{a}}_{n}) &\stackrel{a.s.}{\to}& \bm{C}(\bm{a}).
\end{eqnarray*}
{A second application of the} continuous mapping theorem  implies
\begin{eqnarray*}
   \delta_n:=\det(\bm{C}(\bm{\hat{a}}_{n})) &\stackrel{a.s.}{\to}& \det(\bm{C}(\bm{a}))=:\delta,
\end{eqnarray*}
{and thus there is a set $A$, with $P(A)=1$, such that for every $\omega \in A$ one has $\delta_n(\omega)\to \delta$. Since $\delta=\det(\bm{C}(\bm{a}))\neq 0$, for every $\omega\in A$ there exists an $N(\omega)<\infty$, such that if $n>N(\omega)$ then
  \begin{eqnarray*}
     |\delta_n(\omega)-\delta|&<&\delta/4,
  \end{eqnarray*}
 which is equivalent to that
  \begin{eqnarray*}
     \frac{\delta}{2}< \delta_n(\omega)<\frac{3\delta}{4},
  \end{eqnarray*}
which implies that $\omega \in A_n$, where
  \begin{eqnarray*}
  A_n&=&\{\omega: \delta_n(\omega)>\frac{\delta}{2}\}.
  \end{eqnarray*}
Thus we have established that $\omega\in A$ implies that $\omega \in \cap_{n>N(\omega)} A_n$, which by definition of $\liminf A_n$, means that $A\subset \cup_{m=1}^{\infty} \cap_{n=m}^{\infty} A_n=\liminf A_n$. Finally since $P(A)=1$ this implies that $P(\liminf A_n)=1$. We will establish below that (note that (\ref{fmi}) depend on $n$)
\begin{eqnarray}\label{eq:An_implies_solution}
  A_n\subset \{\mbox{ the equations (\ref{fmi}) have a solution}\}=:B_n,
\end{eqnarray}
which proves that the equations (\ref{fmi}) have a solution eventually, almost surely.
}

{To prove (\ref{eq:An_implies_solution}),} recall that the system in (\ref{fmi}) is given by
\begin{eqnarray}\label{fmi1}
\left\{ 
\begin{array}{c}
q_{1}(1 - p_{1}) + q_{2}(1 - p_{2}) + \dots + q_{s} (1 - p_{s}) = \hat{b}^{(1)}_{n}\\ 
q_{1}(1 - p_{1})p_{1} + q_{2}(1 - p_{2})p_{2} + \dots + q_{s} (1 - p_{s})p_{s} = \hat{b}^{(2)}_{n}\\ 
\dots  \\
q_{1}(1 - p_{1})p_{1}^{k-1} + q_{2}(1 - p_{2})p_{2}^{k-1} + \dots + q_{s} (1 - p_{s})p_{s}^{k-1} = \hat{b}^{(k)}_{n}.
\end{array}
\right. 
\end{eqnarray}
Note that (\ref{fmi1}) can be simplified as 
\begin{eqnarray}\label{fmi2}
\left\{ 
\begin{array}{c}
q_{1} + q_{2} + \dots + q_{s} = \hat{a}^{(1)}_{n}\\
q_{1}p_{1} + q_{2}p_{2} + \dots + q_{s}p_{s} = \hat{a}^{(2)}_{n}\\ 
q_{1}p_{1}^{2} + q_{2}p_{2}^{2} + \dots + q_{s}p_{s}^{2} = \hat{a}^{(3)}_{n} \\ 
\dots \\
q_{1}p_{1}^{k} + q_{2}p_{2}^{k} + \dots + q_{s}p_{s}^{k} = \hat{a}^{(k+1)}_{n},
\end{array}
\right. 
\end{eqnarray}
{where the vector $(\hat{a}^{(1)}_{n},\ldots, \hat{a}^{(k+1)}_{n})$ is defined in $(\ref{eq:anhat-def})$.} The system (\ref{fmi2}), for $k=2s-1$, is a Sylvester-Ramanujan system, cf. \cite{lyubich:2004}. We will here use  Ramanujan's solution, which was published in his third paper in the Journal of the Indian Mathematical
Society cf. \cite{Raman}. From the results in \cite{Raman} it follows that if {$\delta_n=\det(\bm{C}(\bm{\hat{a}}_{n})) \neq 0$}, then {a} solution of (\ref{fmi2}) exists, { which proves (\ref{eq:An_implies_solution}). Furthermore, \cite{Raman} also showed that when the solution exists it is} unique up to permutations of the indices $\{1,\ldots,s\}$ and given by $(\bm{y}, \bm{z})$ which are the coefficients in the parametrisation (\ref{phyz}).

Finally, the pairwise difference of the components of the vector $\tilde{\bm{p}}_{n}$ follows from Theorem 2.1 in \cite{lyubich:2004} and its corollary.
\eop

{Theorem \ref{qptild} gives, in addition to an explicit formula for the derivation of the mle, when it exists, the two results that the components in the vector $\tilde{\bm{p}}_{n}$ are all different, which is as we have already noted a necessary condition for the solution to have a physical meaning, as well as the result that all permutations of the pairs in $(\tilde{\bm{q}}_{n}, \tilde{\bm{p}}_{n})$ are mles, and thus an  mle which is merely assumed to lie in ${\mathcal F}$, and thus is not order restricted,  is {\em not} unique.}

Since the solution is invariant under permutation of the indices, we may choose any permutation and work with it. However, we assumed in  (\ref{eq:ordered-wavelengths}) that the  wavelengths $(\mu_{1}, \dots, \mu_{s})$ are increasing (or, equivalently, the true values of the thinning parameters $(p_{1}, \dots, p_{s})$ are increasing, cf. (\ref{tpp})). Therefore, we choose the unique solution $(\tilde{\bm{q}}, \tilde{\bm{p}})$ of (\ref{fmi2}) which satisfies 
\begin{eqnarray}\label{tildpord}
\tilde{p}_{1} > \tilde{p}_{2} > \dots > \tilde{p}_{s}.
\end{eqnarray}
{Note, however the discussion of an order restricted estimation problem that arises naturally in this setting, in Section \ref{sec:motivation}, which would give an alternative to the above choice.} 

\subsection{Asymptotic properties of the {unordered} mle}\label{subsec:asumpt-mle}
Before we obtain the asymptotic distribution of the estimator we prove an auxiliary lemma. Assume that $k=2s-1$.  We may rewrite the system of equations (\ref{fmi2}) as
\begin{eqnarray}\label{fmi3}
\left\{ 
\begin{array}{c}
F_{1}(\bm{q}, \bm{p}, \bm{u}) = 0\\
F_{2}(\bm{q}, \bm{p}, \bm{u}) = 0\\ 
\dots \\
F_{2s}(\bm{q}, \bm{p}, \bm{u})= 0,
\end{array}
\right. 
\end{eqnarray}
with $\bm{u}=\hat{\bm{a}}_n$, where the functions $F_{i}: \mathbb{R}^{3s} \to \mathbb{R}$ are given by
\begin{eqnarray*}
F_{i}(\bm{q}, \bm{p}, \bm{u}) = q_{1}p_{1}^{i-1} + q_{2}p_{2}^{i-1} + \dots + q_{s}p_{s}^{i-1}  -  u_{i},
\end{eqnarray*}
for $i=1, \dots, 2s$. We see  that the system of equations in (\ref{fmi3}) gives an implicit definition of a function $\bm{\psi}: \mathbb{R}^{2s}= \mathbb{R}^{k+1}\to \ \mathbb{R}^{2s}$, { such that
\begin{eqnarray}\label{eq:psi-def}
    \bm{\psi} (\bm{u} )&=&(\bm{q},\bm{p}).
\end{eqnarray}
}Also, recall that from $2^{s}$ solutions we {choose} the one which satisfies the condition in (\ref{tildpord}), i.e. the components of $\bm{p}$ must be decreasing.

The Jacobian matrix for the system (\ref{fmi3})  is then given by
\begin{eqnarray}\label{Jacm}
\bm{J}(\bm{q}, \bm{p}) = 
 \begin{pmatrix}
 1 &  \cdots & 1 & 0 & \cdots & 0  \\
  p_{1}  &   \cdots & p_{s} & q_{1} & \cdots & q_{s} \\
  p_{1}^{2} & \cdots & p_{s}^{2} & 2q_{1}p_{1} &  \cdots & 2q_{s}p_{s} \\
  \vdots  &  \ddots   &   \vdots &  \vdots & \ddots  &  \vdots  \\
  p_{1}^{2s-1} & \cdots & p_{s}^{2s-1} & (2s-1)q_{1}p_{1}^{2s-2} & \cdots & (2s-1)q_{s}p_{s}^{2s-2}
 \end{pmatrix}
\end{eqnarray}
The next lemma shows that the function $\bm{\psi}(\bm{u})$, implicitly defined by the equations (\ref{fmi3}), is differentiable. 
\begin{lemma}\label{contpsi}
Assume that $\bm{u}$ is such that $\det(\bm{C}(\bm{u})) \neq 0$. Then the function $\bm{\psi}$, implicitly defined by (\ref{fmi3}), is continously differentiable at the point $\bm{u}$.
\end{lemma}
\textbf{Proof.}
The statement of the lemma will follow from the implicit function theorem for which we now check the conditions.

First, we note that (\ref{fmi3}) is a  rewriting of (\ref{fmi2}) which is a simplification of (\ref{fmi1}), and that (\ref{fmi1}) is identical to (\ref{fmi}).  Theorem \ref{qptild} says that if $\det(\bm{C}(\bm{u})) \neq 0$ at some $\bm{u}$, then there is a unique pair $(\bm{q}, \bm{p})$ (with $\bm{p}$ satisfying the {order} condition in (\ref{tildpord})) which satisfy (\ref{fmi}). 

Second, the functions $F_{i}(\bm{q}, \bm{p}, \bm{u})$, for $i=1, \dots, 2s$, are continuously differentiable. 

It remains to prove that the Jacobian $\bm{J}$ in (\ref{Jacm}) is a non-singular matrix, i.e. to show that $\det(\bm{J}) \neq 0$. In fact, we note that $\bm{q}$ can be factored out of the determinant, i.e. 
\begin{eqnarray*}
\det(\bm{J}(\bm{q}, \bm{p}))=q_{1}\cdots q_{s} \cdot \det(\bm{W}(\bm{p})),
\end{eqnarray*}
where
\begin{eqnarray}\label{Jacp}
\bm{W}(\bm{p}) = 
 \begin{pmatrix}
 1 &  \cdots & 1 & 0 & \cdots & 0  \\
  p_{1}  &   \cdots & p_{s} & 1 & \cdots & 1 \\
  p_{1}^{2} & \cdots & p_{s}^{2} & 2p_{1} &  \cdots & 2p_{s} \\
  \vdots  &  \ddots   &   \vdots &  \vdots & \ddots  &  \vdots  \\
  p_{1}^{2s-1} & \cdots & p_{s}^{2s-1} & (2s-1)p_{1}^{2s-2} & \cdots & (2s-1)p_{s}^{2s-2}
 \end{pmatrix}.
\end{eqnarray}
We rewrite $\bm{W}(\bm{p})$ on  column matrix form as
\begin{eqnarray}\label{blvandr}
\bm{W}(\bm{p}) = [\bm{w}(p_{1}), \bm{w}(p_{2}), \dots, \bm{w}(p_{s}), \bm{w}^{(1)}(p_{1}), \bm{w}^{(1)}(p_{2}), \dots, \bm{w}^{(1)}(p_{s})],
\end{eqnarray}
where $\bm{w}(p) = (1, p, p^{2}, \dots, p^{2s})^{T}$, and $\bm{w}^{(1)}(p)$ denotes the vector of componentwise first derivatives of the column vector $\bm{w}(p)$.

Consider $\rho(x)=\det(\bm{W}(x, p_{2}, \dots, p_{s}))$, which is a polynomial of order $(4s-4)$ in $x$. Let us show that the multiplicity of the component $p_{2}$ of the root $(\bm{p},\bm{q})$ is equal to $4$. {In fact,} the third derivative  $\rho^{(3)}(x)$ of the polynomial is equal to
\begin{eqnarray*}
\rho^{(3)}(x) &=&\det([\bm{w}(x)^{(3)}, \bm{w}(p_{2}), \dots, \bm{w}(p_{s}), \bm{w}^{(1)}(x), \bm{w}^{(1)}(p_{2}), \dots, \bm{w}^{(1)}(p_{s})])  \\
&+&3\det([\bm{w}(x)^{(2)}, \bm{w}(p_{2}), \dots, \bm{w}(p_{s}), \bm{w}^{(2)}(x), \bm{w}^{(1)}(p_{2}), \dots, \bm{w}^{(1)}(p_{s})])  \\
&+&3\det([\bm{w}(x)^{(1)}, \bm{w}(p_{2}), \dots, \bm{w}(p_{s}), \bm{w}^{(3)}(x), \bm{w}^{(1)}(p_{2}), \dots, \bm{w}^{(1)}(p_{s})])  \\
&+&\det([\bm{w}(x), \bm{w}(p_{2}), \dots, \bm{w}(p_{s}), \bm{w}^{(4)}(x), \bm{w}^{(1)}(p_{2}), \dots, \bm{w}^{(1)}(p_{s})]) .
\end{eqnarray*}
It follows that for $x=p_2$, each term in the right hand side of the above expression contains two equal columns. Therefore, we have proved that $\rho^{(3)}(x) = 0$ at $x=p_{2}$, which implies that the multiplicity of $p_2$ is at least $4$. 

Now, since $\det(\bm{W}(p_{1}, p_{2}, \dots, p_{s}))$ is symmetric (with no sign change, since flipping two of the arguments $p_i,p_j$ means flipping four columns in the matrix at once), then any $p_{i}$, for $i=2, \dots, s$, is also a root of $\rho(x)=\det(\bm{W}(x, p_{2}, \dots, p_{s}))$, and the same argument as above shows that they all have multiplicity at least $4$. Since  $\rho(x)$ has $s-1$ roots, and it is of order $(4s-4)$, the multiplicity of every root is exactly 4. Therefore, we have shown that
\begin{eqnarray*}
\det(\bm{W}(x, p_{2}, \dots, p_{s})) = c\prod_{j=2}^{s}(x - p_{i})^{4}, 
\end{eqnarray*}
where $c$ is a leading coefficient of the polynomial $\rho(x)$. 
Using the symmetry of the determinant, we may replace any of the $p_i$'s with $x$ and study the upcoming polynomial to obtain 
\begin{eqnarray*}
\det(\bm{J}(\bm{q}, \bm{p})) = c_{1} q_{1}\cdots q_{s} \prod_{p_{i} \neq p_{j}}(p_{i} - p_{j})^{4},
\end{eqnarray*}
where $c_{1}$ is a constant.

Thus, we have shown that $\det(\bm{J}) \neq 0$, provided that $p_{i} \neq p_{j}$ for all $i \neq j$. The fact that the unique (up to permutations of indices) solution $(\bm{p},\bm{q})$ to $ (\ref{fmi})$ satisfies $p_{i} \neq p_{j}$ for all $i \neq j$ follows from Theorem 2.1 and its corollary in \cite{lyubich:2004}.
\eop

\begin{theorem}\label{thmmle}
Let $k=2s -1$. Then the mle $(\hat{\bm{q}}_{n}, \hat{\bm{p}}_{n})$ in (\ref{mle1}) is strongly consistent
\begin{eqnarray*}
(\hat{\bm{q}}_{n}, \hat{\bm{p}}_{n})  \stackrel{a.s.}{\to} (\bm{q}, \bm{p}), 
\end{eqnarray*}
and asymptotically normal
\begin{eqnarray*}
 \sqrt{n}((\hat{\bm{q}}_{n}, \hat{\bm{p}}_{n}) - (\bm{q}, \bm{p})) \stackrel{d}{\to}  \mathcal{N}( \mathbf{0}, \bm{\Sigma}^{2} ),          
\end{eqnarray*}
as $n \to \infty$.


\end{theorem}
\textbf{Proof.}
From Lemma \ref{contpsi} it follows that  for $\bm{u}$, such that  $\det(\bm{C}(\bm{u}) \neq 0$, the system of equations in (\ref{fmi3}) gives an implicit definition of a differentiable function $\bm{\psi}: \mathbb{R}^{2s} \to \mathbb{R}^{2s}$ {such that  $(\bm{q}, \bm{p}) = \bm{\psi}(\bm{a})$, i.e. that (\ref{eq:psi-def}) holds}. 

Combining Theorem \ref{qptild} and Lemma \ref{contpsi},  it follows that {$(\tilde{\bm{q}}_{n}, \tilde{\bm{p}}_{n}) $ is the solution to (\ref{fmi}), eventually, almost surely, so that, furthermore, one has
\begin{eqnarray}\label{eq:tilde-representation-psi}
(\tilde{\bm{q}}_{n}, \tilde{\bm{p}}_{n})  &=&\bm{\psi}(\hat{\bm{a}}_{n}),
\end{eqnarray}
eventually, almost surely} 

{Recall that $\hat{\bm{a}}_{n}\stackrel{a.s}{\to} \bm{a}$, cf.\,$(\ref{eq:an-lim-as})$.}   Then, since $ \bm{\psi}$ is a continuous function, using the continuous mapping theorem, we obtain {$ \bm{\psi} (\hat{\bm{a}}_{n}) \stackrel{a.s}{\to} \bm{\psi}  (\bm{a})$}. {Combining this with (\ref{eq:psi-def}) and (\ref{eq:tilde-representation-psi}) implies}
\begin{eqnarray*}
(\tilde{\bm{q}}_{n}, \tilde{\bm{p}}_{n})  \stackrel{a.s.}{\to} (\bm{q}, \bm{p}).
\end{eqnarray*}
Recall that $(\tilde{\bm{q}}_{n}, \tilde{\bm{p}}_{n})$ is equal to the mle $(\hat{\bm{q}}_{n}, \hat{\bm{p}}_{n})$ only when the restrictions  in $\mathcal{F}$ in (\ref{pqconstr}) are satisfied for $(\tilde{\bm{q}}_{n}, \tilde{\bm{p}}_{n})$, which we will prove below. 

Now, let us consider the vector $\hat{\bm{a}}_{n}$, defined in (\ref{eq:anhat-def}). Note that $[\bm{\hat{a}}_{n}]^{(2, 2s)}$ can be written as
\begin{eqnarray*}
[\bm{\hat{a}}_{n}]^{(2, 2s)} = \bm{1} -  \bm{L}\bm{b}_{n},
\end{eqnarray*}
where $\bm{L}$ is a lower triangular $(2s-1)\times(2s-1)$ matrix of ones. Using a central limit theorem one can show that
\begin{eqnarray}\label{anlim}
 \sqrt{n}([\bm{\hat{a}}_{n}]^{(2, 2s)} - [\bm{a}]^{(2, 2s)}) \stackrel{d}{\to}  \mathcal{N}( \mathbf{0}, \bm{\Sigma}_{A}^{2} ),
\end{eqnarray}
as $n\to\infty$, where 
\begin{eqnarray*}
\bm{\Sigma}_{A}^{2} = \bm{L}\bm{\Sigma}_{m}^{2}\bm{L}^{T},
\end{eqnarray*}
with $\bm{\Sigma}_{m}^{2} = diag([\bm{m}]^{(1, 2s - 1)})$. Recall  that the first element of $\bm{\hat{a}}_{n}$ is deterministic and is equal to $1$, cf. (\ref{eq:anhat-def}), and, thus, we do not include it in the limit result (\ref{anlim}).

Let $\bm{\partial}\bm{\psi}(\bm{u})$ be the matrix of partial derivatives of $\psi(\bm{u})$, i.e.
\begin{eqnarray}\label{partpsi}
\bm{\partial}\bm{\psi}(\bm{u}) = 
 \begin{pmatrix}
  \frac{\partial\psi_{1}}{\partial u_{1}}(\bm{u}) &  \frac{\partial\psi_{1}}{\partial u_{2}}(\bm{u}) &  \frac{\partial\psi_{1}}{\partial u_{3}}(\bm{u}) & \cdots &  \frac{\partial\psi_{1}}{\partial u_{2s}}(\bm{u})\\
\frac{\partial\psi_{2}}{\partial u_{1}}(\bm{u}) &  \frac{\partial\psi_{2}}{\partial u_{2}}(\bm{u}) &  \frac{\partial\psi_{2}}{\partial u_{3}}(\bm{u}) & \cdots &  \frac{\partial\psi_{2}}{\partial u_{2s}}(\bm{u}) \\
 \frac{\partial\psi_{3}}{\partial u_{1}}(\bm{u}) &  \frac{\partial\psi_{3}}{\partial u_{2}}(\bm{u}) &  \frac{\partial\psi_{3}}{\partial u_{3}}(\bm{u}) & \cdots &  \frac{\partial\psi_{3}}{\partial u_{2s}}(\bm{u})  \\
  \vdots  & \vdots  & \vdots &  \ddots  & \vdots  \\
\frac{\partial\psi_{2s}}{\partial u_{1}}(\bm{u}) &  \frac{\partial\psi_{2s}}{\partial u_{2}}(\bm{u}) &  \frac{\partial\psi_{2s}}{\partial u_{3}}(\bm{u}) & \cdots &  \frac{\partial\psi_{2s}}{\partial u_{2s}}(\bm{u})
 \end{pmatrix}.
\end{eqnarray}
The values of  $\bm{\partial}\bm{\psi}(\bm{u})$ can be found using the implicit function theorem. In fact, the $j$-th column $ \bm{\partial}\bm{\psi}(\bm{u})[j]$ of $\bm{\partial}\bm{\psi}(\bm{u})$ is the solution of the following system of linear equations
\begin{eqnarray*}
\bm{J} \bm{\partial}\bm{\psi}(\bm{u})[j] = \bm{1}^{(j)},
\end{eqnarray*}
where $\bm{1}^{(j)}\in \mathbb{R}^{2s}$ is defined by $\bm{1}^{(j)}_j = -1$ and $\bm{1}^{(j)}_l = 0$ for $l \neq j$, cf. (\ref{fmi3}) and (\ref{Jacm}). The solution exists, and it is unique, when $\det(\bm{J}) \neq 0$, which is true for $\bm{u}=\bm{\hat{a}}_{n}$ for all $n\geq n_1$ and for $\bm{u}=\bm{a}$.  Thus, the matrices $\bm{\partial}\bm{\psi}(\bm{\hat{a}}_{n})$ are (uniquely) given for all $n\geq n_1$, and so is the matrix $\bm{\partial}\bm{\psi}(\bm{{a}})$.

Since the derivatives $\bm{\partial}\bm{\psi}$ are continuous, using (\ref{anlim}) and the delta method we derive the limit distribution for $(\tilde{\bm{q}}_{n}, \tilde{\bm{p}}_{n})$
\begin{eqnarray*}
 \sqrt{n}((\tilde{\bm{q}}_{n}, \tilde{\bm{p}}_{n}) - (\bm{q}, \bm{p})) \stackrel{d}{\to}  \mathcal{N}( \mathbf{0}, \bm{\Sigma}^{2} ),          
\end{eqnarray*}
as $n \to \infty$, with 
\begin{eqnarray*}
\bm{\Sigma}^{2} = [\bm{\partial}\bm{\psi}(\bm{u})]_{1:2s, 2:2s}\times\bm{\Sigma}_{A}^{2}\times[\bm{\partial}\bm{\psi}(\bm{u})]_{1:2s, 2:2s}^{T},
\end{eqnarray*}
where  {$[\cdot]_{1:2s, 2:2s}$  denotes} a matrix without the first column.

Finally, $(\hat{\bm{q}}_{n}, \hat{\bm{p}}_{n}) = (\tilde{\bm{q}}_{n}, \tilde{\bm{p}}_{n})$ if and only if $(\tilde{\bm{q}}_{n}, \tilde{\bm{p}}_{n}) \in \mathcal{F}$, with $\mathcal{F}$ defined in (\ref{pqconstr}). Since  $(\tilde{\bm{q}}_{n}, \tilde{\bm{p}}_{n})$ is strongly consistent, there exists $n_{2} > n_{1}$ such that for all $n > n_{2}$ 
\begin{eqnarray*}
\mathbb{P}[(\tilde{\bm{q}}_{n}, \tilde{\bm{p}}_{n}) \in \mathcal{F}] = 1.
\end{eqnarray*}
Thus, the mle $(\hat{\bm{q}}_{n}, \hat{\bm{p}}_{n})$ is consistent, almost surely, and has the same asymptotic distribution as $(\tilde{\bm{q}}_{n}, \tilde{\bm{p}}_{n})$, which ends the proof.
\eop

We note that having obtained an estimator of $\bm{p}$, one can use a functional relation between a thinning probability and a wavelength value, cf. \cite{willis:carlile:1999}, similarly to as in Corollaries 1 and 2 in \cite{anevskistocpr}. The derivation is straightforward and is omitted.

\section{Order restricted estimation of the parameters}\label{sec:order-restricted}
We note that the components in the mle are not necessarily ordered vectors, and that we have order restrictions on both the wavelength distribution $\bm{q}$ and the thinning probabilities $\bm{p}$. 

We therefore address order restricted problems in this section. {A natural approach would be to use any permutation $(\tilde{\bm{q}}_{n}, \tilde{\bm{p}}_{n})$ of the solution to the mle in Theorem \ref{qptild}, as a  preliminary, starting estimator,  and modify it in some way, in order to obtain an order restricted estimator. One approach for this would be to do isotonic regression for the partial order $\succeq$ introduced in $(\ref{eq:partial-order})$. This could potentially yield a reasonable estimator, using recent results in \cite{anevski:pastukhov:2018-general} on limit distributions for isotonic regression with respect to a partial order. However, as noted in the discussion in Section \ref{sec:motivation}, the parameter space defined by imposing the partial order $\succeq$ is not closed, which makes the optimisation problem nonstandard. We therefore refrain from using that approach.}

{Instead we use a two step approach consisting of first estimating the unknown order from the order of mle of $\bm{p}$, which by assumption is ordered with strict inequalities, and second of projecting the correspondingly ordered mle of $\bm{q}$ on that estimated order.}  

{Thus, let $(\tilde{\bm{q}}_n, \tilde{\bm{p}}_n)$ be any permutation of the solution of (\ref{fmi2}). Define 
\begin{eqnarray}\label{eq:two-step.1}
     (\bar{\bm{p}}_n,\hat{\xi}_n)&=&sort(\tilde{\bm{p}}_n)
\end{eqnarray}
as the ordered (in decreasing order) vector $\bar{\bm{p}}_n$ of $\tilde{\bm{p}}_n$, and its index (rank) $\hat{\xi}_n$, so that $\bar{\bm{p}}_n=\tilde{\bm{p}}_{n,\hat{\xi}_n}$, where we use the notation $ x_{\xi}= (x_{\xi}(1),\ldots, x_{\xi}(s))=(x(\xi(1)),\ldots,x(\xi(s)))$ for any permutation $\xi$. Let $\bar{\bm{q}}_n=\tilde{\bm{q}}_{n,\hat{\xi}_n}$ be the corresponding mle of $\bm{q}$. This is the preliminary estimator of $\bm{q}$ arising from the first step in the approach that we now suggest. In the second step we define the final order restricted estimator $\hat{\bm{q}}$ as the $l^2$-projection of  $\bar{\bm{q}}$ on on the set of decreasing probability mass functions.}

{The treatment in the sequel is slightly reversed compared to the two step procedure proposed above. Thus in Subsection \ref{subsec:decreasing} we first} treat order restricted inference for the wavelength distribution $\bm{q}$, {using an $l^2$ projection on the space of decreasing pmf's, when we are given a preliminary estimator that has a known limit distribution}. {Next, in Subsection  \ref{subsec:decreasing2} we combine the derived result with a result on the almost sure consistency of the estimated order  $\hat{\xi}_n$ as well as with the limit distribution of the unrestricted mle of $\bm{q}$ derived in Theorem \ref{thmmle}, to derive the limit distribution for the final two step procedure estimator.}

\subsection{Estimating a decreasing wavelength distribution, for known index order}\label{subsec:decreasing}
In this subsection we assume that it is known that the wavelength distribution $\bm{q}$ is a decreasing vector, and construct an appropriate estimator, based on the mle defined previously. In fact, our estimator is the $l^2$-projection of the mle of $\bm{q}$ on the space of positive and decreasing vectors, i.e. the isotonic regression of the vector $\hat{\bm{q}}_n$. {Note that in this subsection we assume that the index order in which  $\bm{q}$  is decreasing is known. This can alternatively and equivalently be  stated as the assumption that the version of the mle $\hat{\bm{q}}_n$ that we use has the correct index order, although the correspoding mle $\hat{\bm{q}}_n$ itself may be not ordered.}

We define the set $\mathcal{Q}^{*} \subset \mathbb{R}^{s}$ 
\begin{eqnarray}\label{pqconstrdecr} 
\mathcal{Q}^{*}  = \{\bm{q}\in  \mathbb{R}^{s}:   q_{1} \geq q_{2} \geq  \dots  \geq q_{s} \}
\end{eqnarray}
and assume that the true value satisfies $\bm{q} \in \mathcal{Q}^{*}$. Note first that since $\bm{q}$ is supposed to be a probability mass function, we should really demand that  $\mathcal{Q}^{*}$ is a subset of positive $s$-dimensional vectors and furthermore that there should be a linear constraint. {The linear coinstraint} is however not necessary when projecting a vector that already is a probability mass function, since isotonic regression preserves linear constraints as well as upper and lower bounds of the vector, cf. \cite{robertsonorder} for these results and a general overview of order restricted inference.

We define the monotone constrained estimator of $\bm{q}$ as
\begin{eqnarray}\label{decrest} 
\hat{\bm{q}}^{*}_{n} = \underset{q \in \mathcal{Q}^{*} }{\arg \min} \sum_{r =1}^{s} (q_{r} - \hat{q}_{n,r})^2,
\end{eqnarray}
i.e. $\hat{\bm{q}}^{*}_{n}$ is the isotonic regression of the mle $\hat{\bm{q}}_{n}$.
We note that from the error reduction property of the isotonic regression we have
\begin{eqnarray}\label{erred} 
||\hat{\bm{q}}^{*}_{n} - \bm{q} ||_{\alpha} \leq ||\hat{\bm{q}}_{n} - \bm{q} ||_{\alpha} 
\end{eqnarray}
for all $\alpha \geq 1$, cf. \cite{robertsonorder}.

In order to obtain the asymptotic distribution of $\hat{\bm{q}}^{*}_{n}$, we need to specify the exact shape of the pmf $\bm{q}$, since the shape of $\bm{q}$ will determine the limit distribution.  In particular we need to specify the regions where $\bm{q}$ is constant. Thus we assume that the true vector $\bm{q}\in\mathbb{R}^{s}$ has the following structure
\begin{eqnarray}\label{restflat}
	q_{t_{1}} = \dots = q_{t_{1} + v_{1} -1} > q_{t_{2}} &=& \dots = q_{t_{2} + v_{2} -1} > \dots >\\\nonumber
	q_{t_{m}} &=& \dots = q_{s},
\end{eqnarray}
where $t_{j}$ for $j = 1, \dots m$ is the index of the first element in the $j$-th flat region, $q_{t_{1}} = q_{1}$, $m$ is the total number of flat regions of $\bm{q}$, $\bm{v} = (v_{1}, \dots, v_{m})$ is the vector of the lengths (the numbers of points) of the flat regions of $\bm{q}$, so that $\sum_{j=1}^{m}v_{j}=s$.

We define the map $\eta=\eta_{\bf{q}}: \mathbb{R}^{s} \to \mathbb{R}^{s}$ by specifying that for  any $Y \in \mathbb{R}^{s}$, for all constant regions $(t_{j},t_{j} + v_{j} -1)$ of $\bm{q}$,
\begin{eqnarray}\label{eq:eta-def}
[\eta(Y)]^{(t_{j},t_{j} + v_{j} -1)} = \arg \min_{\bm{y}\in \{\bm{y}\in {\mathbb R}^{v_j}:y_1\geq \ldots\geq y_{v_{j}}\} }||Y^{(t_{j},t_{j} + v_{j} -1)}-\bm{y}||^2,
\end{eqnarray}
where $||\cdot||^{2}$ denotes the $l^2$-norm in $ {\mathbb R}^{v_j}$, so that the values of $[\eta(Y)]^{(t_{j},t_{j} + v_{j} -1)}$ are given as the separate isotonic regression of $Y$ over the region of constancy $(t_{j},t_{j} + v_{j} -1)$. Note that if the region of constancy is of length one then the isotonic regression of $Y$ at that region (point) is equal to the value of $Y$ at that point.  With this definition, we see that  $\eta(Y)$ is a concatenation of separate isotonic regressions over each region of constancy of the true $\bm{q}$, cf. \cite{jankowski:wellner:2009} and \cite{anevski:pastukhov:2018-general} for a more detailed description of the map (operator).

Finally we obtain consistency and the asymptotic distribution of the estimator $\hat{\bm{q}}^{*}_{n}$.
\begin{theorem}\label{thm:order-mle}
Suppose $\bm{q}$ satisfies (\ref{restflat}), and let $k=2s-1$. Then the order restricted estimator  $\hat{\bm{q}}^{*}_{n}$ defined in (\ref{decrest}) is strongly consistent
\begin{eqnarray*}
\hat{\bm{q}}^{*}_{n}  \stackrel{a.s.}{\to} \bm{q}, 
\end{eqnarray*}
and has the asymptotic distribution
\begin{eqnarray*}
 \sqrt{n}(\hat{\bm{q}}^{*}_{n} - \bm{q}) \stackrel{d}{\to} \eta(\bm{Q_{\bm{q}}}),          
\end{eqnarray*}
as $n \to \infty$, where $\bm{Q}_{\bm{q}}$ is the limit distribution of $\hat{\bm{q}}_{n}$, i.e. $\bm{Q}_{\bm{q}} = \mathcal{N}( \mathbf{0}, [\bm{\Sigma}^{2}]_{1:s, 1:s})$, with $\bm{\Sigma}^{2}$ defined in Theorem \ref{thmmle}, { and $\eta$ is the map defined in (\ref{eq:eta-def}).}
\end{theorem}
\textbf{Proof.}
The strong consistency follows from the consistency of the mle  $\hat{\bm{q}}_n$ and the error reduction property of the isotonic regression. The asymptotic distribution of $\hat{\bm{q}}^{*}_{n}$ follows by Theorem 2 in \cite{anevski:pastukhov:2018-general}, see also Theorem 5.2.1 in \cite{robertsonorder}, and \cite{jankowski:wellner:2009}.
\eop

\subsection{Estimating a decreasing wavelength distribution, for unknown index order}\label{subsec:decreasing2}
{
In this subsection we state a limit distribution for the two step procedure proposed in the beginning of this section. 
}

{
We first introduce some notation. Note that by assumption, $\bm{p}$ is strictly ordered.  In the definition (\ref{eq:two-step.1}) of the ordered mle of $\bm{p}$, the mle will be an estimator of the unknown vector of probabilities $\bm{p}$, in some unknown index permutation, i.e. the sorted mle defined in (\ref{eq:two-step.1}) will be an estimator of $\bm{p}_{\xi}$, for some unknown permutation $\xi$ of the set $\{1,2,\ldots,s\}$.}

{
Now let $\hat{\xi}_n$ be the estimated index order, defined in (\ref{eq:two-step.1}).  Then, since (the unordered) mle $\hat{\bm{p}}_{n}$ is almost surely consistent by Theorem \ref{thmmle}, then by the continuous mapping theorem
\begin{eqnarray*}
     \hat{\xi}_n &\stackrel{a.s.}{\to}& \xi,
\end{eqnarray*}
as $n\to\infty$.}

{
Further, recall that $\bar{\bm{q}}_n=\tilde{\bm{q}}_{\hat{\xi},n}$ is the starting estimator, obtained as the mle of $\bm{q}$ corresponding to the sorted mle of $\bm{p}$, cf. (\ref{eq:two-step.1}).  We can now define the final estimator, obtained as the $l^2$ projection of $\bar{\bm{q}}_n$ on the space of decreasing pmfs,
\begin{eqnarray}\label{decrest} 
\hat{\bm{q}}^{\hat{\xi}}_n= \underset{q \in \mathcal{Q}^{*} }{\arg \min} \sum_{r =1}^{s} (q_{r} - \bar{q}_{n,r})^2,
\end{eqnarray}
i.e. $\hat{\bm{q}}^{\hat{\xi}}_{n}$ is the isotonic regression of the mle $\bar{\bm{q}}_{n}$. }

{
\begin{theorem}\label{thm:order-mle-estim}
Suppose $\bm{q}$ satisfies (\ref{restflat}), let $\hat{\bm{q}}^{\hat{\xi}}_n$ be defined in $(\ref{decrest} )$,  and let $k=2s-1$. Then $\hat{\bm{q}}^{\hat{\xi}}_n$ is strongly consistent
\begin{eqnarray*}
\hat{\bm{q}}^{\hat{\xi}}_n  \stackrel{a.s.}{\to} \bm{q}, 
\end{eqnarray*}
and has the asymptotic distribution
\begin{eqnarray*}
 \sqrt{n}(\hat{\bm{q}}^{\hat{\xi}}_n - \bm{q}) \stackrel{d}{\to} \eta(\bm{Q_{\bm{q}}}),          
\end{eqnarray*}
as $n \to \infty$, where $\bm{Q}_{\bm{q}}$ is the limit distribution of $\hat{\bm{q}}_{n}$, i.e. $\bm{Q}_{\bm{q}} = \mathcal{N}( \mathbf{0}, [\bm{\Sigma}^{2}]_{1:s, 1:s})$, with $\bm{\Sigma}^{2}$ defined in Theorem \ref{thmmle}, { and $\eta$ is the map defined in (\ref{eq:eta-def}).}
\end{theorem}
\textbf{Proof.} The conclusion of theorem is a straightforward consequence of the result $ \hat{\xi}_n \stackrel{a.s.}{\to} \xi$ and Theorem \ref{thm:order-mle}.
\eop}

\section{Discussion and conclusions}\label{sec:discussion}
In this paper we have derived the mle $(\hat{\bm{q}}_n,\hat{\bm{p}}_n)$ of the distribution of events of different types $\bm{q}$, i.e. marks, of a homogeneous marked Poisson process and the thinning probabilities $\bm{p}$, based on data from sequential thinning of a Poisson process. We have established that the number, $k$, of sequential thinnings needed in order to solve a system of algebraic equations that determines the mle is $k=2s-1$, where $s$ is the length of the vector $\bm{q}$, cf. Theorem \ref{qptild}. In Theorem \ref{thmmle} we derived the strong consistency and asymptotic normality of the mle $(\hat{\bm{q}}_n,\hat{\bm{p}}_n)$. 

{We would like to emphasise that in the assumptions for the experiment, that we have performed, we state that although the values of the wavelengths are assumed to be unknown, we however a priori may know their order, and this is given in (\ref{eq:ordered-wavelengths}). Thus the indices $1,2,\ldots,s$ correspond to an ordered set of wavelengths and one goal in this paper has been to estimate their distribution $\bm{q}$. An estimator of $q$ is then given as $\hat{\bm{q}}^*_n$. A possibly reasonable loosening of the model assumptions in a real world physics experiment may be to assume that the order of the wavelengths is {\em unknown}. One may assume that there {\em is} an order (\ref{eq:ordered-wavelengths}) for the unknown wavelengths, but that one does not know the indices, or labels,  $1,2,\ldots,s$ that one uses to give the correct ordering. Thus, the problem would be to estimate $\bm{q}$ under the assumption of an order on the values of  $\bm{q}$ (which is ordered in the reverse way to the wavelengths) but in which one does not know the correct order. An estimator of $q$ is then given as  $\hat{\bm{q}}_n^{\hat{\xi}_n}$. A problem which is reminiscent to this later approach, was treated in \cite{anevski:gill:zohren:2017}, in which one derived a likelihood based estimator for an unknown ordered probability mass function in which one does not know the correct order. It may be of interest to attempt to adapt the method in \cite{anevski:gill:zohren:2017} to the problem treated in this paper. One should note that in \cite{anevski:gill:zohren:2017} however, one allowed for there being more than the observed $s$ "species" (corresponding to wavelengths in our setting), and also that the observed wavelengths may be different than the $s$ most probable wavelengths, and that one thereby derived a estimator of the {\em whole} wavelength distribution, allowing for infinitely many possible wavelengths. Thus the setting in \cite{anevski:gill:zohren:2017}  is not immediately applicable to our problem, although it is related.}

We have constructed order {restricted estimators $\hat{\bm{q}}^*_n$ and $\hat{\bm{q}}^{\hat{\xi}}_n$ of $\bm{q}$, and in Theorems \ref{thm:order-mle} and \ref{thm:order-mle-estim} we derived the consistency and asymptotic distribution of $\hat{\bm{q}}^*_n$ and $\hat{\bm{q}}^{\hat{\xi}}_n$, respectively. The estimator  $\hat{\bm{q}}^*_n$ is the mle derived under the assumption that the order is known. The estimator $\hat{\bm{q}}^{\hat{\xi}}_n$ is a two-step estimator derived under the assumption of unknown order, where one first derives a strongly consistent estimator $\hat{\xi}_n$ of the unknown order and next uses that to derive the order restricted mle $\hat{\bm{q}}_n^{\hat{\xi}_n}$, under the estimated order.}

{We have also considered estimating the vectors $\bm{p}$ and $\bm{q}$ simultaneously, and have considered an order restricted estimator under a partial order (\ref{eq:partial-order}). We refer to the last paragraphs of Section \ref{sec:motivation} for a discussion of challenges for this approach.}

{Finally we would like to note that a} possible way to improve the efficiency for the order restricted estimator may be to use model selection to choose the appropriate class of probability mass functions $\bm{q}$. The model class may be determined by the regions of constancy of $\bm{q}$, as defined (\ref{restflat}). One advantage with having knowledge about the specific sets of regions of constancy for the unknown vector is that one can then use the knowledge to construct an order restricted estimator that outperforms the regular isotonic regression estimator, as shown in  \cite{anevski:pastukhov:models-selection:2018} . In \cite{anevski:pastukhov:models-selection:2018} we introduced an information criterion which can be used for model selection in order restricted inference and also we have provided a post model selection estimator, and derived asymptotic properties for it. An attempt to adapt the methods developed in  \cite{anevski:pastukhov:models-selection:2018} to the problem treated in this paper may be of interest.






\begin{acks}[Acknowledgments]
VP's research was fully supported by the Swedish Research Council (SRC), and the research of DA was partially supported by the SRC. The
authors gratefully acknowledge the SRC's support. We would also like to thank Victor Ufnarovski and Andrey Ghulchak for their kind help with Lemma 3.
\end{acks}

\end{document}